\newif\ifsubmode
\submodefalse

\newif\ifprintfig
\printfigtrue

\newif\ifemulate
\emulatetrue
\ifsubmode
\documentclass[12pt,preprint]{aastex}
\received{}
\accepted{}
\journalid{}
\articleid{}
\else
   \documentclass{emulateapj}
   \submitted{{\it To be submitted for publication in ApJ}}
\fi
\usepackage{multirow}

\newcommand{\Msun}{\mbox{\,$M_{\odot}$}}

\def\spose#1{\hbox to 0pt{#1\hss}}
\def\simlt{\mathrel{\spose{\lower 3pt\hbox{$\mathchar"218$}}
     \raise 2.0pt\hbox{$\mathchar"13C$}}}
\def\simgt{\mathrel{\spose{\lower 3pt\hbox{$\mathchar"218$}}
     \raise 2.0pt\hbox{$\mathchar"13E$}}}

\shorttitle{Chemical Abundances of Stellar Halos}
\shortauthors{Zolotov~et~al.}

\begin{document} 


\title{The Dual Origin of Stellar Halos II: Chemical Abundances as Tracers of Formation History}
\author{Adi\ Zolotov\altaffilmark{1}, Beth\ Willman\altaffilmark{2},
  Alyson\ M.\ Brooks\altaffilmark{3}, Fabio\
  Governato\altaffilmark{4}, David\ W.\ Hogg\altaffilmark{1}, Sijing\
  Shen\altaffilmark{5}, James\ Wadsley\altaffilmark{5}}

\altaffiltext{1}{Center for Cosmology and Particle Physics, Department of Physics, New York University, 4 Washington Place, New York, NY 10003; az481@nyu.edu}
\altaffiltext{2}{Haverford College, Department of Astronomy, 370 Lancaster Avenue, Haverford, PA 19041; bwillman@haverford.edu}
\altaffiltext{3}{California Institute of Technology, M/C 350-17, Pasadena, CA 91125}
\altaffiltext{4}{Department of Astronomy,University of Washington, Box 351580, Seattle, WA 98195}
\altaffiltext{5}{Department of Physics and Astronomy, McMaster University, Hamilton, Ontario, L88 4M1, Canada}

\date{June 28, 2010}

\begin{abstract}
Fully cosmological, high resolution N-Body + SPH simulations are used
to investigate the chemical abundance trends of stars in simulated
stellar halos as a function of their origin. These simulations employ
a physically motivated supernova feedback recipe, as well as metal
enrichment, metal cooling and metal diffusion. As presented in an
earlier paper, the simulated galaxies in this study are surrounded by
stellar halos whose inner regions contain both stars accreted from
satellite galaxies and stars formed in situ in the central regions of
the main galaxies and later displaced by mergers into their inner
halos.  The abundance patterns ([Fe/H] and [O/Fe]) of halo stars
located within 10 kpc of a solar-like observer are analyzed. We find
that for galaxies which have not experienced a recent major merger, in
situ stars at the high [Fe/H] end of the metallicity distribution
function are more [$\alpha$/Fe]-rich than accreted stars at similar
[Fe/H]. This dichotomy in the [O/Fe] of halo stars at a given
[Fe/H] results from the different potential wells within which in
situ and accreted halo stars form. These results qualitatively match
recent observations of local Milky Way halo stars. It may thus be
possible for observers to uncover the relative contribution of
different physical processes to the formation of stellar halos by
observing such trends in the halo populations of the Milky Way, and
other local $L^{\star}$ galaxies.
\end{abstract}

	  
\section{INTRODUCTION}\label{intro_sec}

Observational evidence has been mounting that the stellar halo of the
Milky Way (MW) was likely assembled through a combination of both
hierarchical accretions and in situ star formation. The properties of
local halo stars suggest that the Galaxy's halo is composed of at
least two distinct but overlapping stellar populations
\citep{Carollo2010}. These two halo components exhibit different
spatial distributions, orbits and metallicities, most likely because
each formed through a different mechanism, and on a different
timescale. While the outer halo component, which dominates at
distances greater than $\sim 20$ kpc, with [Fe/H] $\sim -2.5$, is
consistent with having been accreted from merged dwarf galaxies, the
stars in the more metal rich ([Fe/H] $\sim -1.5$) inner halo likely
formed predominantly in situ
\citep{Norris1994,Chiba2000,Carollo2007, Carollo2010}.

Studies of the detailed chemical abundances of MW halo stars have
sought to place further constraints on the formation of the Galactic
halo
\citep{Nissen1997,Stephens2002,Venn2004,Pritzl2005,Ishigaki2009}. 
These works rely on the behavior of the [$\alpha$/Fe] ratio to
indicate the timescale of the star formation history of the
halo. Alpha elements are primarily produced in core-collapse
supernovae, which occur after $\sim 10^7$ years, while the majority of
iron is produced in supernovae Type Ia, whose timescale of $\leq 10^9$
years is significantly longer
\citep{Maoz2010}.  
In the metallicity range of $-2<[Fe/H]<-0.5$, inner halo stars are
$\sim 0.1$ dex more alpha rich than outer halo stars, implying that
the Galaxy's halo was not uniformly enriched by core-collapse
supernovae \citep{Ishigaki2009}. In a local sample of stars, those
with [$\alpha$/Fe] $\sim 0.3$ show a constant [$\alpha$/Fe] over the
above metallicity range, while stars with lower $\alpha$-ratio, of
about 0.15, exhibit a decrease in [$\alpha$/Fe] toward higher
metallicity
\citep{Nissen2010}. 
Similar declines to lower [$\alpha$/Fe] have been observed in dwarf
galaxies, though at lower [Fe/H] values \citep{Tolstoy2009}. These trends
strengthen the argument that the Milky Way's halo likely hosts two
populations of stars, with distinct formation origins.

In the past year, theoretical results from galaxy simulations have
been used to investigate the contribution of in situ stars to the
formation history of stellar halos. Using fully cosmological N-Body +
Smooth Particle Hydrodynamic (SPH) simulations of four stellar halos
surrounding disk galaxies
\citet[][hereafter Paper I]{Zolotov2009} showed that while the formation of
 most stellar halos is dominated by accretion events, a population
of in situ stars resides in the inner regions of halos as well.  These
in situ halo stars form in the inner most regions of the simulated galaxy
and are displaced into the stellar halo through
major mergers. In situ stars may also end up in stellar halos, along
side accreted stars, through minor accretion events, which heat up
disks and eject a small fraction of their stars into kinematic halo
orbits \citep{Purcell2009}.

Previous work on the chemical abundance trends of stellar halos has
relied on simulations that only allowed for the formation of stellar
halos via hierarchical buildup, using dark-matter only simulations,
combined with semi-analytic models \citep{Bullock2005,Robertson2005,
Font2006a,Font2006c,Johnston2008, DeLucia2008}. These simulations,
however, do not have the capability to examine possible chemical
abundance trends for in situ halo stars.

In this paper, we use four high resolution cosmological SPH+N-Body
simulations of disk galaxies to investigate the chemical abundance
trends of both accreted and in situ halo stars. While past numerical
work on chemical abundance trends in halo stars have focused solely
on accreted stars, the goal of this work is to test whether in situ
and accreted stars are distinguishable using chemical abundance
patterns.

\section{Simulations}
The four high-resolution disk galaxy simulations used in this study
were chosen to have approximately the same total mass, though with
varied merging histories. Three of these simulations (MW1,
h277, and h285) are discussed in detail in Paper I and references
therein, while the remaining simulation (h258) is described in
\citet{Governato2009}. These disk dominated galaxies were simulated using
 GASOLINE \citep{Wadsley2004}. We briefly summarize the important
features of these simulations below. MW1 has a total mass of $10^{12}
M_{\odot}$, while the remaining three galaxies, h277, h258, and h285,
all have a total mass of $\sim 7.6 \times 10^{11} M_{\odot}$. All have
several million particles (dark matter + gas + stars) within their
virial radius at z=0. Table 1 lists the primary properties of all four
simulations. A physically motivated recipe was used to describe star
formation and supernova feedback \citep{Stinson2006, Governato2007}
with a uniform UV background that turns on at $z=9$, resembling cosmic
reionization \citep{Haardt2005}.

Because this work aims to examine the metallicity and alpha abundances
 of halo stars, the details of the treatment of chemical enrichment
 are important to this paper. All of the galaxies used in this study
 have been rerun since Paper I with metal cooling and metal diffusion,
 which realistically captures the effect of the turbulent interstellar
 medium (ISM) on metal mixing \citep{Shen2009}. Metal cooling
 increases the amount of star formation in these galaxies, and
 therefore the amount of energy deposited into the local ISM by
 supernova has been increased from 0.4, in Paper I, to 0.7 $\times
 10^{51}$ ergs. The supernova feedback employed in these simulations
 deposits the energy into the ISM, and cooling is turned off for a
 period of time for those gas particles within the supernova blast
 wave radius. As described in detail in \citet{Stinson2006}, supernova
 Type Ia and II yields are adopted from \citet{Thielemann1986}, and
 \citet{Woosley1995}, respectively, and implemented following
 \citet{Raiteri1996}. We follow oxygen and iron yields from SNe type
 Ia and core-collapse SNe, and use [O/Fe] as our proxy for
 [$\alpha$/Fe]. These galaxies follow observed local metallicity
 trends \citep{Brooks2007}, as well as metallicity trends at high
 redshift \citep{Maiolino2008,Pontzen2008,Pontzen2010}.

\subsection{Selecting ``Observed'' Halo Stars}\label{sec_halo}

Each simulated galaxy used in this study has been decomposed into a
disk, bulge, and halo kinematic component, using the full
three-dimensional phase-space information available for all stars. We
use the same decomposition technique as in Paper I, but provide all
the relevant details here. In order to decompose the disk from the
spheroid, we first align the angular momentum vector of the disk with
the z-axis. This places the disk in the x-y plane, and allows us to
calculate the angular momentum of each star in that plane, $J_z$, as
well as the momentum of the co-rotating circular orbit with similar
orbital energy, $J_{circ}$. A star with a circular orbit in the plane
of the disk will have $J_z/J_{circ} \sim 1$ in this framework. Stars
with $J_z/J_{circ} \geq 0.8$ are selected as disk stars. This cut is
equivalent to an eccentricity cut of $e \leq 0.2$, which matches the
eccentricities observed in the Milky Way's disk
\citep{Nordstrom2004}. 

In this study, we use only those stars which have orbital
characteristics of a stellar halo population. The spheroid (halo +
bulge) of each galaxy is defined using a cut in $J_z/J_{circ}$ such
that the spheroidal population does not exhibit a net rotation. We can
further decompose the spheroid into a bulge and halo because a break
in the mass density profile is observed. Stars which are tightly bound
to the galaxy - those whose total energy is low- are classified as
bulge stars, while all other stars are identified as halo stars. The
energy cut employed is set in order to make a two-component fit to the
mass profile of the spheroid. Bulge stars are within the break of the
mass profile, and halo stars are outside the break. While there are
stars whose kinematics do not make either the disk or spheroid cuts
(thick disk and pseudo bulge stars), they are not relevant to this
study. Only stars which have been identified by this procedure as
belonging to the kinematic halo population are used in this study.

After we have selected the kinematic halo population using the above
procedure, we go on to create an observational sample similar to the
sample of halo stars in SDSS used by \citet{Carollo2010}.  To simulate
an observer at the solar position, we place the ``Sun'' at a distance
of $\sim$ 3 disk radial scale lengths away from the center, and then
calculate the heliocentric distances of each star. The radial scale
length of each simulated disk is listed in Table 1. While
\citet{Carollo2010} only observe stars out to a volume less than 4 kpc
from the sun, we have chosen to study halo stars within 10 kpc of the
simulated heliocentric observer. The foremost reason for this 10 kpc
cut-off is that nearly all of the in situ stars ($\geq 80 \%$,
described in the following section) in the simulated halos are within
the inner $\sim 10$ kpc of the stellar halos. Second, many studies of
the Milky Way's stellar halo have used F \& G type stars, which are
currently observed out to a few tens kpc. To match a large ground
based survey like SDSS we restrict our observed sample to stars more than 30
degrees above and more than 1 kpc above the plane of the disk, for
a heliocentric observer.

Stars that are classified as belonging to the halo in the kinematic
decomposition, and which fit the spatial and SDSS-like cuts are
referred to as part of the observed halo sample. Table 2 lists the
total stellar mass in each galaxy's halo, as well as the fraction of
in situ and accreted stars in each observed halo sample.

\begin{deluxetable*}{lcccccc}
\tabletypesize{\scriptsize}
\tablecaption{PROPERTIES OF SIMULATED GALAXIES}
\tablewidth{0pt}
\tablehead{
\colhead{Run} &
\colhead{$M_{vir}$}&
\colhead{$N_{tot}$ at z=0} &
\colhead{$M_{particle}^{DM}$} &
\colhead{$M_{particle}^{\ast}$} &
\colhead{$\epsilon$\tablenotemark{a}}&
\colhead{Radial Scale Length}\\
\colhead{} &
\colhead{$\Msun$} &
\colhead{dark+gas+stars}&
\colhead{$\Msun$} &
\colhead{$\Msun$} &
\colhead{kpc} &
\colhead{kpc} 
}
\startdata
MW1hr&$1.1 \times 10^{12}$& $ 4.9 \times 10^6$& $ 7.6 \times 10^5$&$ 2.7\times 10^4$&0.3 & 2.0\\
h277&$7.4 \times 10^{11}$&$ 2.3 \times 10^6$&$ 1.2 \times 10^6$ &$ 4.6 \times 10^4$&0.35& 2.6\\
h285&$7.7 \times 10^{11}$&$ 3.0 \times 10^6$&$ 1.2 \times 10^6$&$ 4.6 \times 10^4$&0.35&2.9\\
h258&$7.2 \times 10^{11}$&$ 2.5 \times 10^6$&$ 1.2 \times 10^6$&$ 4.6 \times 10^4$&0.35 & 4.0\\
\enddata
\tablenotetext{a}{Gravitational spline softening length}

\end{deluxetable*}

\subsection{Halo Star Origin - Accreted vs. in situ}\label{ssec_origin}

We trace the formation history of each star particle located within
the virial radius of the main galaxy at $z=0$ back to $z=6$, as well
as follow the gas particles from which the stars have formed. At each
time step output by the simulation (every 25 Myr), we identify the
dark matter halo to which each particle belonged using
AHF\footnotemark \citep{Gill2004,Knollmann2009}. A star is considered
bound to a dark matter halo only if it is identified as belonging to
that halo for at least two consecutive time steps. Using this
technique we identify three different formation origins: accreted, in
situ, and ambiguous. For a detailed description of this procedure, and
a discussion of numerical issues, the reader is referred to Paper
I. We quickly review the three classifications below.

\footnotetext{AMIGA's HALO FINDER, available for download at http://popia.ft.uam.es/AMIGA}

Accreted stars formed in halos other than the main galaxy's dark
matter halo. Through merging, these stars have become unbound to their
progenitors and now belong to the main galaxy's halo. The majority of
the accreted stars in the halos of h277, h285, and MW1 (75\% by mass
in stars), originated in $2-3$ subhalos, each with a total mass of
$1-3 \times 10^{10} M_{\odot}$. The majority of accreted stars in h258,
which experiences several major mergers, as discussed in Section 3,
originate in three galaxies, with total masses ranging from 2.5 to 6
$\times 10^{10} M_{\odot}$.

In situ stars, on the other hand, formed within the main galaxy's
potential well, with their gas progenitors bound to the main galaxy
before their formation. Paper I found this population formed within
the inner $\sim 4$ kpc of the galaxy's center before being displaced
into the kinematic halo component as a result of mergers. Stars which
are classified as ambiguous, $ \sim 3 \%$ of the total halo stars,
have an unknown formation history due to the limited number of time
steps output for each simulation. This occurs if a gas particle spawns
a star particle at approximately the same time that it first becomes
bound to the primary. In such cases it is uncertain whether the star
formed in the primary or in its original subhalo.

All four of the simulated galaxies analyzed in this paper host a
stellar halo which contains both accreted and in situ stars. While the
relative contribution of each population differs from halo to halo,
the presence of stars with a dual origin is a generic feature of the
stellar halos surrounding $L^{\star}$ galaxies at z=0. In Paper I we
showed that the fraction of in situ stars present in a halo depends
strongly on the merging history of the primary galaxy. The in situ
population of galaxies with very active merging histories, and which
therefore host massive accreted halos, is diluted in comparison to the
in situ halo population of galaxies which have not accreted as many
stars during their lifetimes. As suggested by their in situ fractions,
h285 and h258 have much more active merging histories than MW1 and
h277 (Table 2).

\begin{deluxetable*}{llccccc}
\tabletypesize{\scriptsize}
\tablecaption{STELLAR HALO ORIGINS}
\tablewidth{0pt}
\tablehead{
\colhead{Run} &
\colhead{total halo mass}&
\colhead{accreted fraction} &
\colhead{in situ fraction} &
\colhead{ambiguous fraction\tablenotemark{a}} &
\colhead{$z_{lmm}$\tablenotemark{b}}\\
\colhead{} &
\colhead{$M_{\odot}$}&
\colhead{\% in obs sample} &
\colhead{\% in obs sample} &
\colhead{\% in obs sample} &
\colhead{}
}
\startdata
MW1hr &$6.9 \times 10 ^9$  &65 & 31 & 4 & 3\\

H277 &$9.22\times 10^9$ & 72 & 25 & 3 & 3 \\

H285 &$2.7\times 10^{10}$ & 91 & 6 & 3 & 2 \\

H258 &$1.59 \times 10 ^{10}$ & 87 & 10 & 3 & 1 \\
\enddata 
\tablecomments{
The total mass listed for each halo is the total stellar mass within
the virial radius. The accreted, in situ, and ambiguous fractions are
for the stars within each observed halo sample, as described in Section 2.1}

\tablenotetext{a}{The origin of these stars is ambiguous due
to the finite time resolution of the simulation, as described in
Section 2.2}
\tablenotetext{b}{The redshift of last significant major merger, defined as a merger with mass ratio $\frac{M_{primary}}{M_{secondary}} < 7$.}

\end{deluxetable*}

\section{TRENDS IN CHEMICAL ABUNDANCES WITH FORMATION ORIGIN}
We now examine the trends in chemical abundances displayed by the
observed halo stars.  One expects that the chemical abundance pattern
of a stellar population will be set by the star formation history of
the galaxy in which the population formed. For this reason we aim to
find trends which differentiate the in situ and accreted populations
in each stellar halo. These qualitative trends in metallicity are
robust, and applicable not just to the Milky Way, but to Milky Way-mass
galaxies with different merging histories, as discussed in detail
below, and in Section 4.

Massive galaxies form in deep potential wells, and consequently have
higher star formation rates (SFR) than less massive galaxies, resulting
in a mass-metallicity relation
\citep{Tremonti2004,Brooks2007,Finlator2008}. With their high SFR,
massive galaxies will reach higher [Fe/H] at relatively constant
[$\alpha$/Fe] before SNe Ia begin to contribute, as core-collapse SNe
enrich the ISM with mostly alpha elements, and little iron or
iron-peak elements. In lower mass galaxies, however, lower SFRs mean
that the galaxy can't enrich as much in Fe before SNe type Ia began to
contribute, resulting in a decrease of [$\alpha$/Fe] at lower
[Fe/H]. Additionally, galaxies with long star formation histories
will form stars out of gas that has been enriched by both
core-collapse SNe and SNe Type Ia, whereas galaxies with truncated
star formation histories will have their abundance pattern set
primarily by core-collapse SNe. These trends are shown in Figure 1.

If the in situ and accreted populations each formed at different
times, under different physical conditions, we expect to see that
reflected in their abundance patterns. We indeed find that in three of
our galaxies (those without a very recent major merger) in situ halo
stars at the high [Fe/H] end of the metallicity distribution function
(MDF) tend to be more alpha rich than the similarly high [Fe/H]
accreted stars (Figure 3, discussed in detail below). At lower [Fe/H],
we find that the two populations have similar [O/Fe]. Because we only
follow oxygen and iron in these simulations [O/Fe] serves as our proxy
for [$\alpha$/Fe].

Before we discuss our findings in detail, it is important to first
note that the satellites of the simulated galaxies studied here are
brighter, and hence are more [Fe/H]-rich, than those observed in the
Milky Way \citep{Zolotov2009}. If the satellites which built up our
simulated stellar halos contained less stellar mass at a given total
mass the qualitative chemical abundance patterns for in situ and
accreted halo stars discussed below would not change, as we
demonstrate in detail in Section 4. The results in metallicity we
discuss, however, are not meant to be absolute, and the qualitative
trends are robust for Milky Way-mass galaxies.

As discussed in Section 2.2, in situ stars formed in the innermost
regions of the primary galaxy's dark-matter halo, and were displaced
into the halo through massive mergers. Since the redshift of the last
significant merger (Table 2) determines the last time at which in situ
stars are substantially displaced into the halo, the formation time of
the in situ populations of each galaxy is quite different. The left
panel of Figure 2 shows the time of formation for in situ halo stars
(solid lines) and accreted halo stars (dashed lines) in each simulated
observed sample.  Because MW1 and h277 experience their last
significant merger early on, at z$\sim 3$, the in situ stars in these
galaxies all formed before $t<2.5$ Gyr. The late mergers of h285 and
h258, at z=2, and z=1, respectively, result in younger in situ
populations in these halos. The right panel of Figure 2 shows the time
at which accreted halo stars first became unbound from their
satellites and were accreted onto the primary halo. While the majority
of the accreted stars in h277 and MW1 become bound to these galaxies
early on (with 90 \% of such stars being accreted at times $\leq 4$
Gyr), h285 and h258 accreted a large number of their halo stars at
times later than this.

Figure 3 shows [Fe/H] vs [O/Fe] of in situ halo stars (in red
triangles) and accreted halo stars (in black circles) for each
simulated observed halo sample. We find that in MW1, and h277 (top two
panels of Figure 3), where $80\%$ of in situ halo stars formed within
a 1 Gyr time span in a deep potential well, the ISM did not have time
to enrich with SNe Type Ia, and so the [$\alpha$/Fe] of these stars is
approximately constant with [Fe/H]. The accreted stars in these halos,
however, which formed slightly later on, and in shallower potential
wells than the in situ stars, experienced a slower star formation
rate, resulting in a decrease of their [$\alpha$/Fe].

While the accreted stars of h285 (bottom left panel of Figure 3)
display a similar trend in [$\alpha$/Fe] to the accreted populations
of h277 and MW1, its in situ population does not display a constant
[$\alpha$/Fe] like the other galaxies. The in situ population of this
galaxy had more time to form (because of the later merger), and formed
in a smaller galaxy than MW1 and h277's in situ stars. At z=3, the total virial
masses of h277 and MW1 are $8.5
\times 10^{10} M_{\odot}$ and $1 \times 10^{11} M_{\odot}$,
respectively, whereas h285's mass is only $5 \times 10^{10}
M_{\odot}$. Because of this, the in situ stars of h285 display a turn
over in [O/Fe] at lower [Fe/H], rather than remaining constant at
higher metallicities like MW1 and h277.

Unlike the other galaxies, h258 (bottom right panel of Figure 3)
experiences a nearly 1:1 merger at z=1. The in situ and accreted stars
in this galaxy do not follow a similar trend to the other
galaxies. The in situ halo population in this galaxy had longer to
form (Figure 2, left panel), and hence were enriched to a lower
[$\alpha$/Fe] than the other in situ halo populations. Furthermore,
because of its binary merger, many of the h258 in situ stars have
similar chemical abundances to the accreted population. This galaxy
has undergone three massive accretion events in its lifetime, whose
``tracks'' are evident in Figure 3.

\begin{figure*}[t!]
\epsscale{0.7}
\plotone{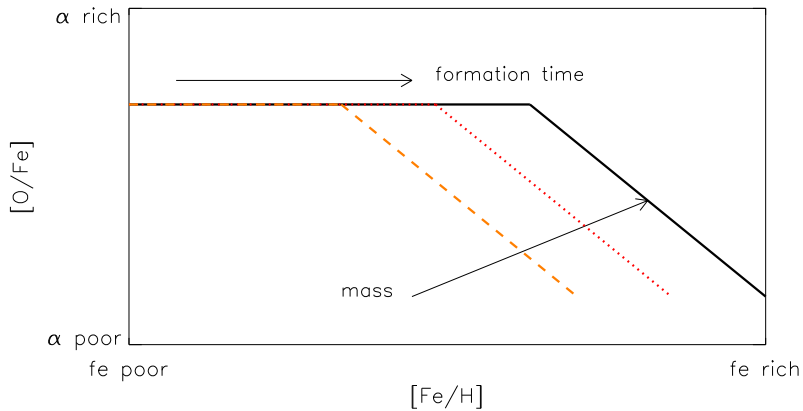}
\caption{
This cartoon shows the evolution of [Fe/H] vs [$\alpha$/Fe] (tracked
by [O/Fe] in our simulations) as a function of a stellar population's
formation time, and the mass of the galaxy in which they form. The
black solid line represents the most massive galaxy, the red dotted
line an intermediate mass galaxy, and the orange dashed line a low
mass galaxy. Higher mass galaxies have higher star formation rates,
and therefore enrich with more core-collapse supernova in a given time,
reaching a higher [Fe/H] before turning over to lower [$\alpha$/Fe]
compared to a low mass galaxy with a slower SFR in the same amount of
time.}
\end{figure*}

\begin{figure*}[t!]
\epsscale{0.8}
\plottwo{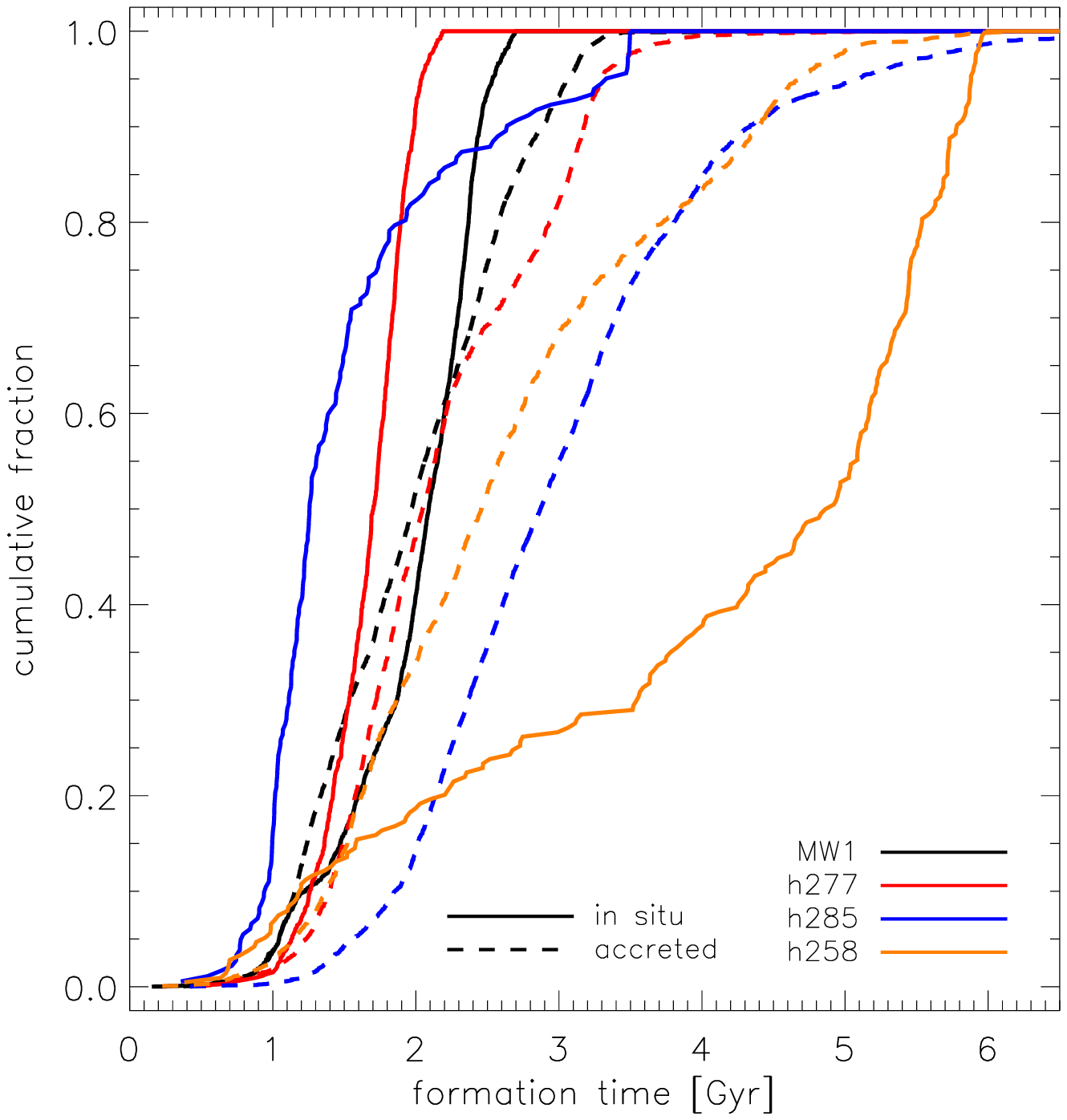}{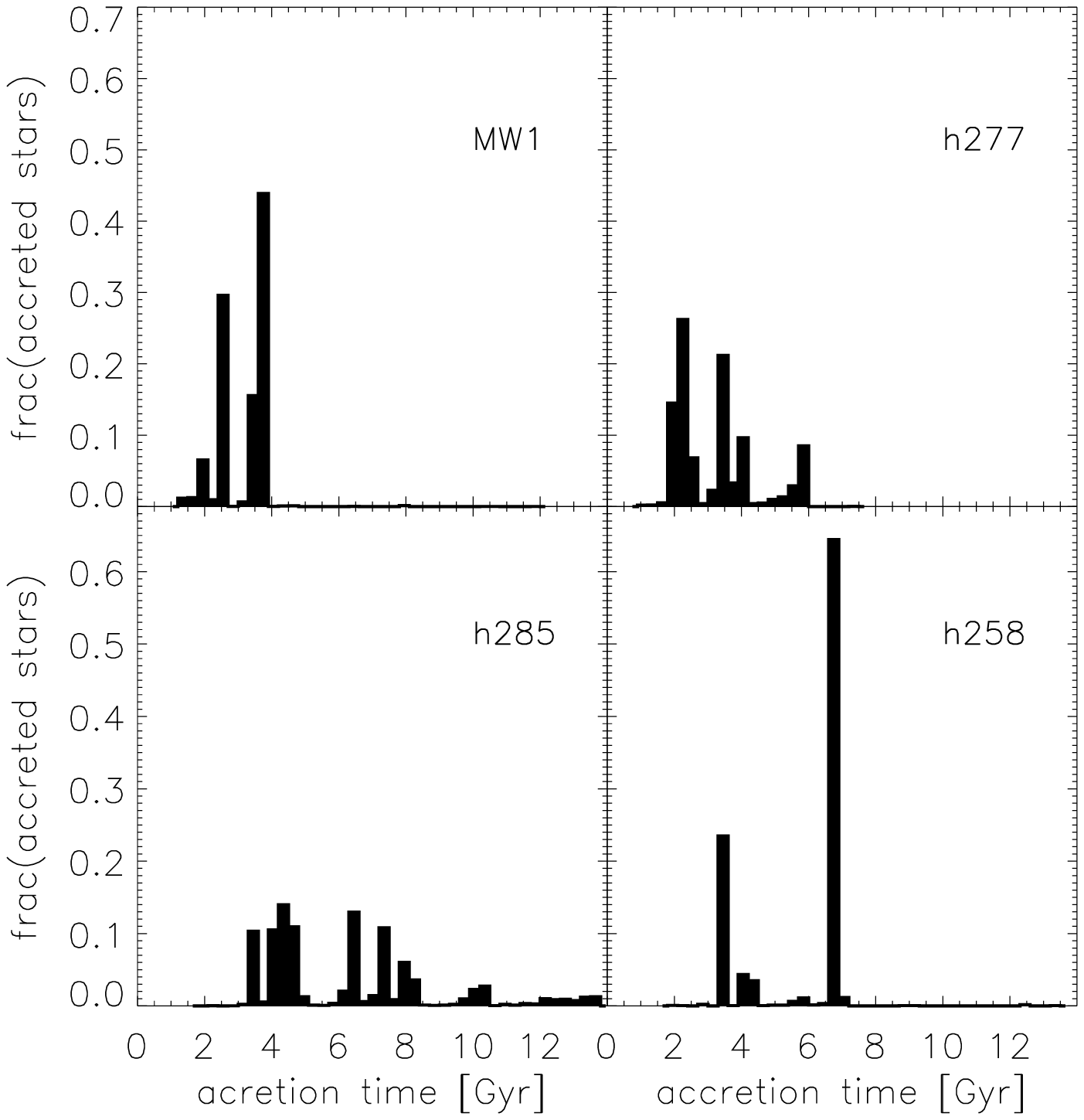}
\caption{
Left Panel: The fractional contribution of accreted and in situ stars
to each of the four simulated observed halo samples, as a
function of the populations' formation time. In situ stars are shown
in solid lines, while accreted stars are shown in dashed lines.  Right
Panel: The accretion history of each simulated observed halo sample. The
accretion time shown is the time at which each star became unbound
from its progenitor and was accreted onto the primary halo. Both h285
and h258 have a higher contribution from late accretion events than
MW1 and h277. }
\end{figure*}

\begin{figure*}[t!]
\epsscale{1.0}
\plottwo{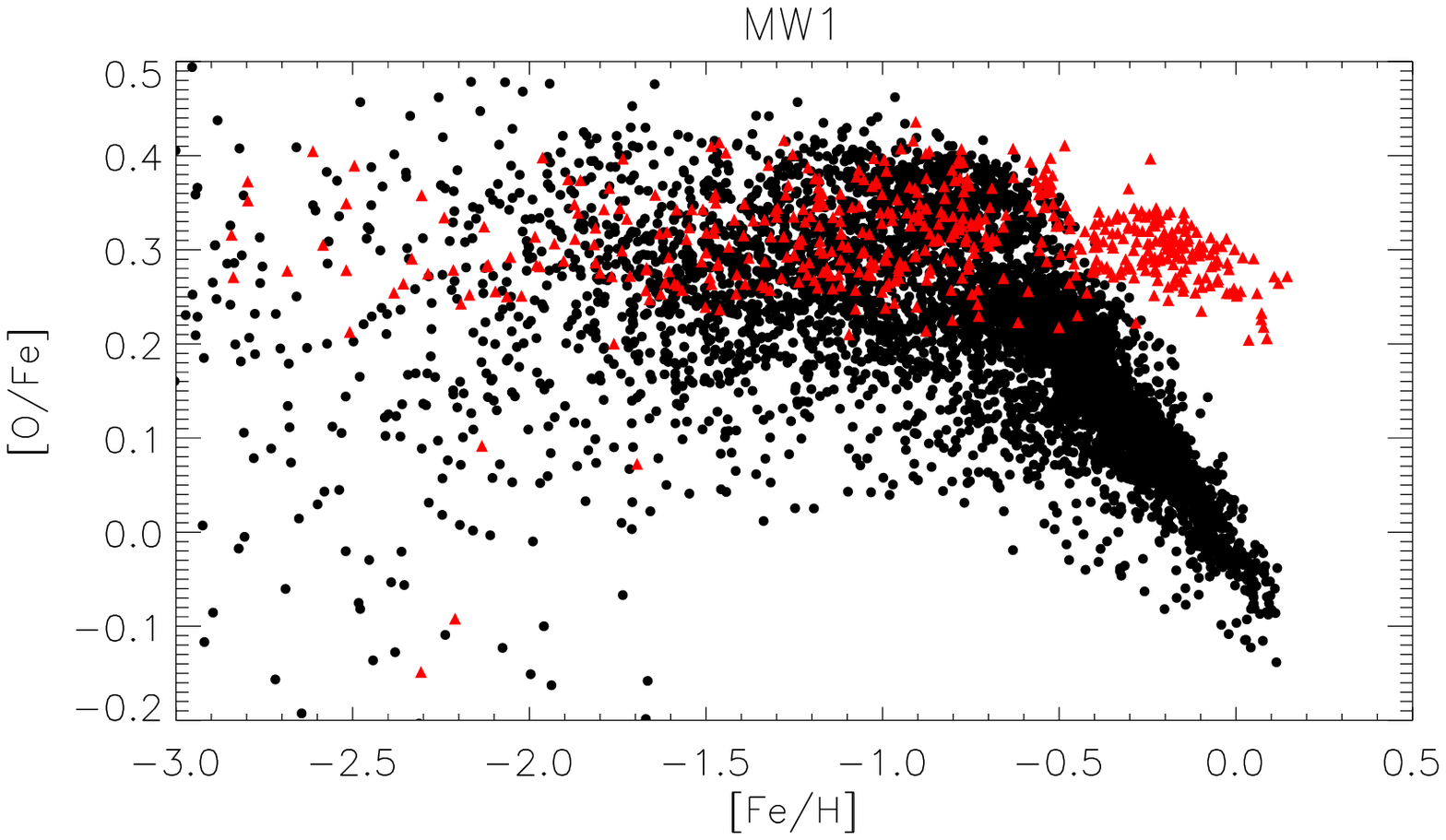}{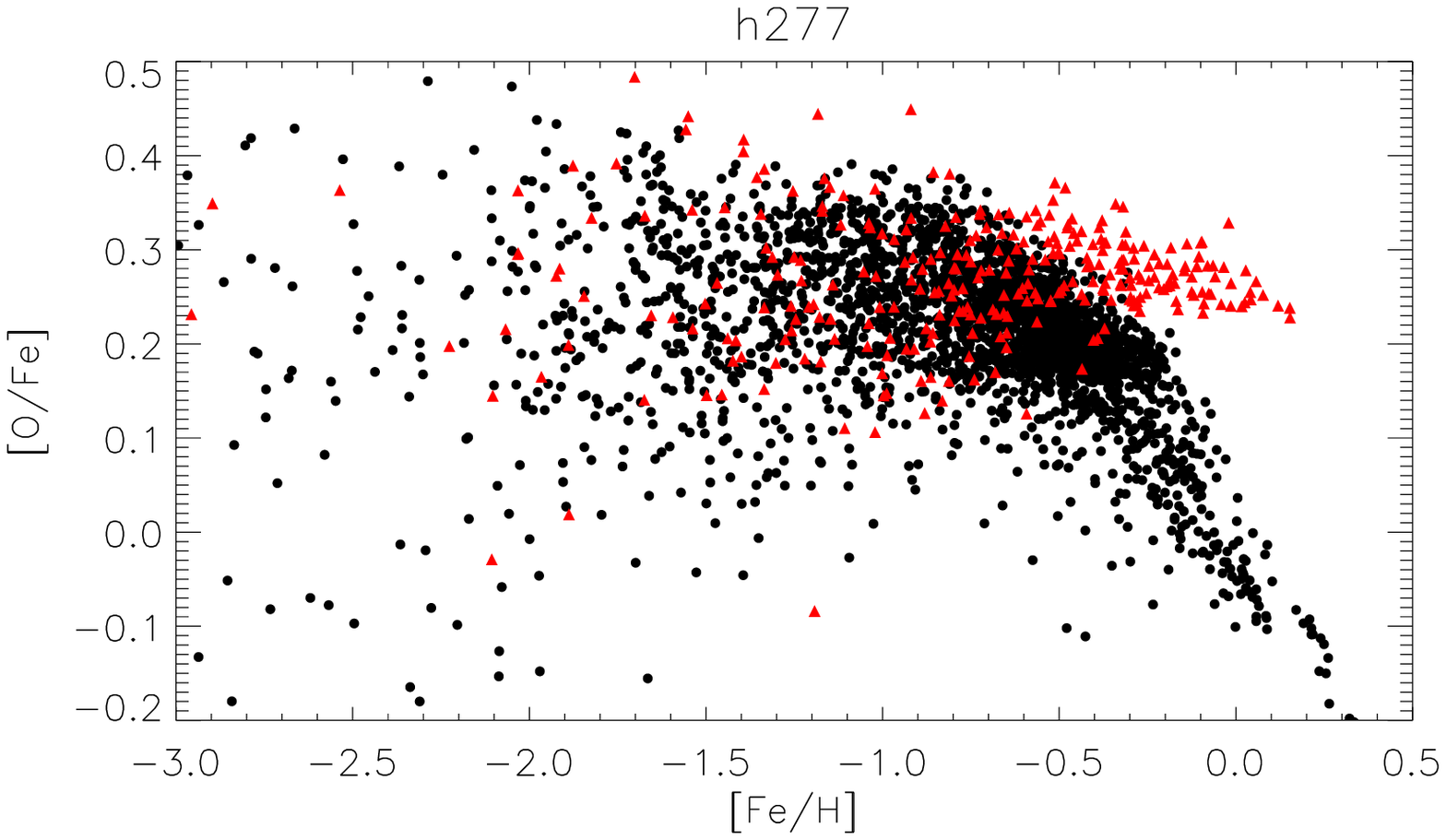}
\epsscale{1.0}
\plottwo{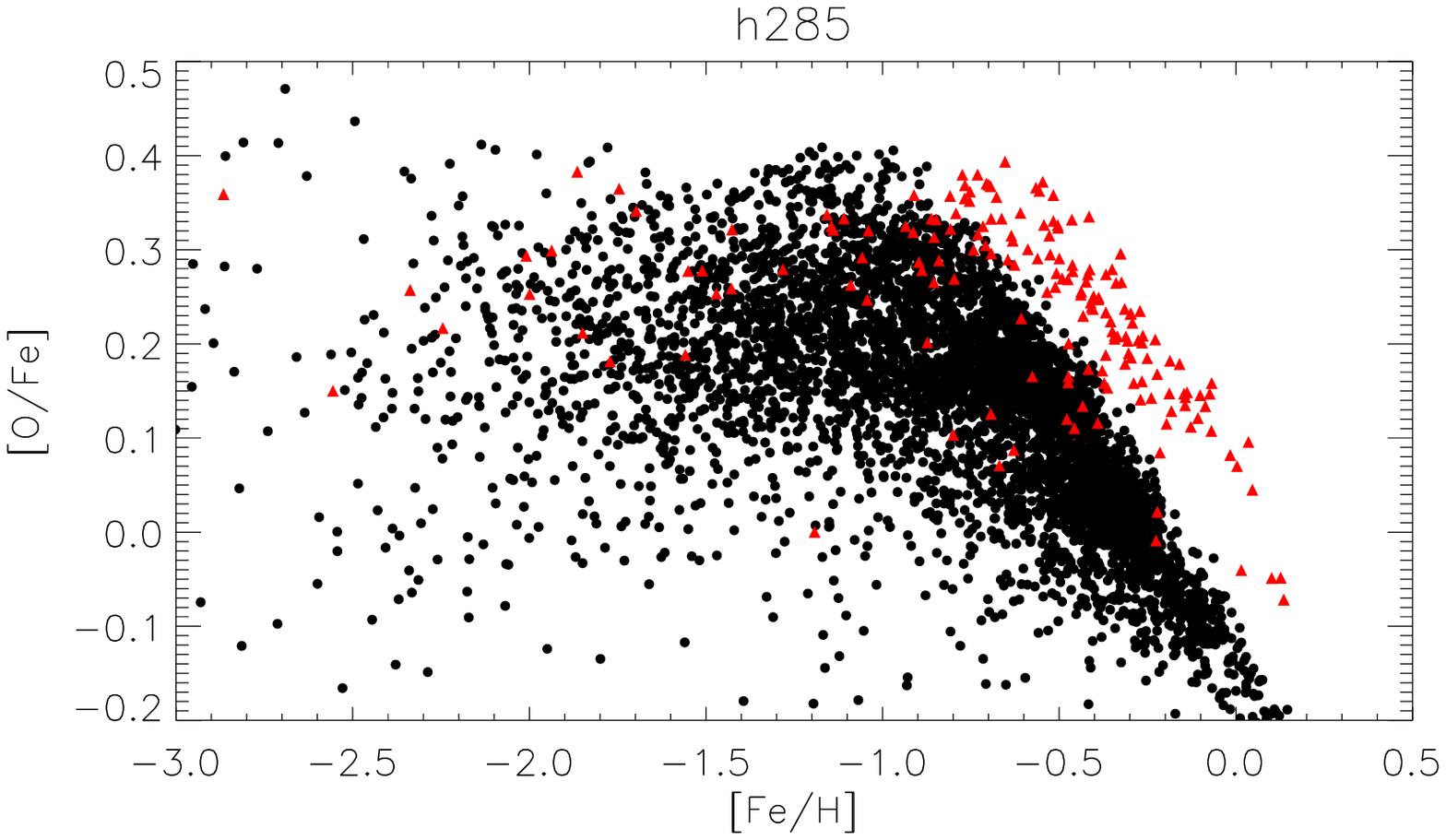}{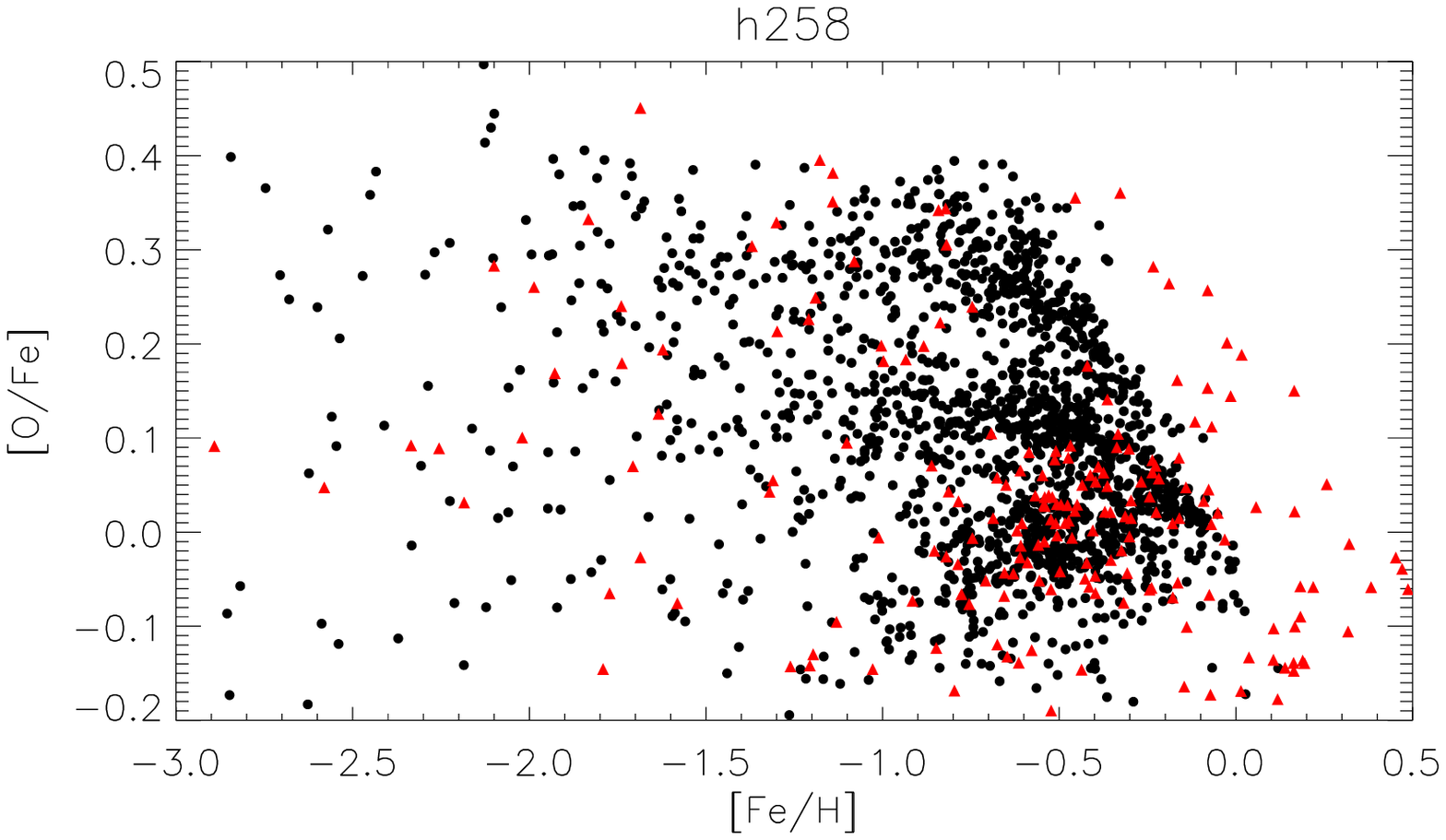}
\caption{
[O/Fe] vs [Fe/H] for observed halo stars in each of the four simulated
galaxies studied in this paper. In situ stars are shown in red
triangles, and accreted stars in black dots. Accreted stars dominate
the stellar halos of all the galaxies studied here. The in situ stars
of MW1, h277, and h285, which formed relatively early on in the center
of these galaxies, are more alpha-rich at high metallicity than
accreted stars. The same is not true for h258 (shown in the bottom
right panel), as discussed in Section 3. The satellites of the
galaxies studied here are overluminious, and so halo stars are more
metal-rich in this study than expected from observations, as discussed
in detail in Section 4.
\label{fig_radial}}
\end{figure*}

\section{Simulation Effects}
The primary result of this study is that in situ halo stars have
higher [$\alpha$/Fe] than accreted halo stars at the high [Fe/H] end
of the MDF. In this section, we discuss the role of simulation
effects in this study and the robustness of our primary results to
these effects. 

The most relevant numerical issue in this study is that the satellites
of the simulated galaxies are brighter than those observed in the
Milky Way
\citep{Zolotov2009}. While the Milky Way hosts only two
satellites with $-19 < M_v < -16$, the Large and Small Magellanic
Clouds, the four simulated galaxy halos have, on average, four
satellite companions in this luminosity range at z=0. Brighter
satellites will of course contribute more accreted stars to a given
stellar halo.  This affects our results in two pertinent ways. First,
the fractional contribution of in situ stars to each stellar halo is
likely under-estimated, as discussed in Paper I. Second, because our
simulated satellites contain too many stars, they are also more [Fe/H]
rich than we expect for the satellites that contributed to Milky
Way-mass stellar halos.

The differing [$\alpha$/Fe] trends displayed by the in situ and
accreted halo components is a robust prediction for Milky Way-massed
galaxy halos that is not undermined by the details of the present
simulations' MDFs. Had our simulated halos been built up from
satellites that were less luminous at a given total satellite mass,
the trends shown in Figure 3 would have simply been shifted to lower
[Fe/H]. To demonstrate this, we have re-simulated one of the galaxies
in this paper (MW1) with the same exact initial conditions but with
differently calibrated supernova feedback and star formation. The
physical treatment of these processes is unchanged in the new
simulation. The new run employed a stronger supernova feedback, with
an increase in thermal SNe energy from 0.7 to 1 $\times 10^{51}$ ergs,
and the threshold for star formation density has been increased to
from 0.1 to 1 amu/$cm^{3}$.  As a result of the stronger feedback, and
higher density threshold, at z=0 the new simulated halo hosts only two
satellites in the luminosity range $-19< M_v <-16$.  The new MW1
stellar halo also has a MDF that is similar to that observed of halo
stars in M31 \citep{Kalirai2006}. In the inner 30 kpc of its stellar
halo, stars in the new MW1 have $<[Fe/H]> \sim -0.9$.

Figure 4 shows [O/Fe] vs [Fe/H] for a z=2 satellite for the new run
with stronger feedback (top panel), and for the MW1 used in this paper
(bottom panel). We have chosen to look at z=2 satellites because
$95\%$ of accreted stars which end up in our halo sample (see Figure
2) formed before this time in such satellites. The total mass of both
satellites is $\sim 9.4 \times 10^9 M_{\odot}$, but the stellar mass
of the new satellite is $4.5 \times 10^8$, and of the old satellite
is $1.5 \times 10^9 M_{\odot}$. Both satellites exhibit the
characteristic turn-over to lower [O/Fe]. The new run with the less
luminous satellite population simply exhibits this turn-over at a
lower [Fe/H]. The in situ stars of each galaxy are shown in red
triangles. Both simulations show that in situ stars are more
alpha-rich at the high [Fe/H] end than the satellite stars. The
distinct [$\alpha$/Fe] trends of the two halo populations is a result
of their formation in potential wells whose depths vary by more than
an order of magnitude. The trends do not depend on the details of the
simulations' MDF.

\begin{figure*}[t!]
\epsscale{0.6}
\plotone{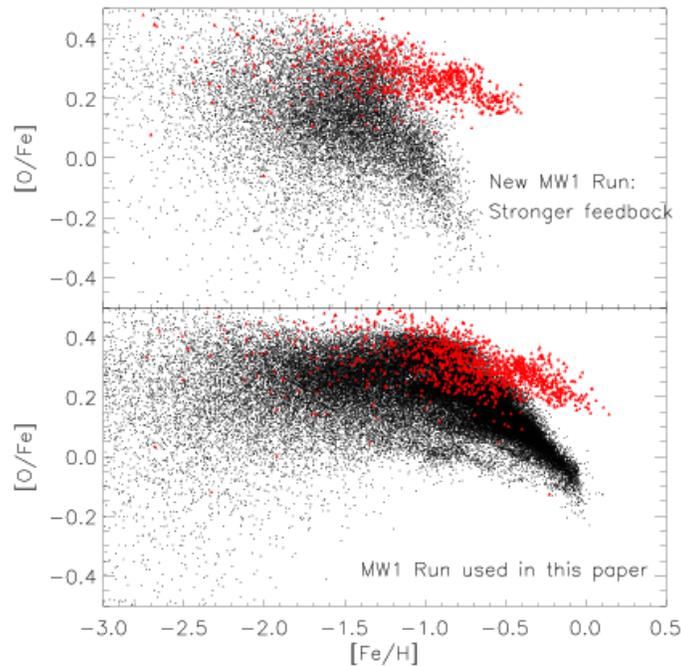}
\caption{
[O/Fe] vs [Fe/H] for z=2 satellite stars (shown in black) in one of
the simulations discussed in this paper (bottom panel) and in a
satellite of a new simulation, with identical initial conditions, but
a stronger SNe feedback (top panel). In situ stars are shown in red
triangles. See section 4 for details. Such satellites are typical
contributors to the z=0 accreted populations in our halos. Employing a
stronger SNe feedback in the simulations results in more metal-poor
stars, though both runs display the same qualitative trend that the in
situ stars are more alpha rich than accreted stars at the high
metallicity end. }

\end{figure*}

\section{DISCUSSION \& CONCLUSIONS}

In this paper, we have studied the chemical abundance trends of inner
halo stars in four SPH+N-Body simulations of approximately L* disk
galaxies. We investigate the possibility of using chemical abundance
trends to discern the relative importance of in situ and accretion
processes in building up the inner stellar halos of Milky Way-mass
galaxies.  For each simulated galaxy we concentrated on a sample of
halo stars located within 10 kpc of a simulated sun-like
observer. This sample included stars from two populations, one formed
in situ, and one accreted. We find that in the inner halos of galaxies
where a recent binary merger did not occur, high [Fe/H] in situ halo
stars are more [$\alpha$/Fe] rich than accreted stars at similar
[Fe/H].

The bimodal distribution of [$\alpha$/Fe] at high [Fe/H] for in situ
and accreted stars is primarily due to the different mass galaxies in
which they formed. The in situ halo populations (in all but h258) were
formed early on deep in the potential wells of the primary dark matter
halos, where a high star formation rate ensured that only
core-collapse SNe dominated the chemical enrichment up to high
[Fe/H]. The accreted stars, which formed later on average than the in
situ stars, and in shallower potential wells, underwent a slower
chemical evolution, where SNe Type Ia contributed iron at lower
[Fe/H], resulting in a lower [$\alpha$/Fe] at a given [Fe/H]. Such
trends are expected given a mass-metallicity relation
\citep{Brooks2007}. Though the satellite galaxies in our simulations
are too bright, and hence too metal rich, we have found that the
relative chemical abundance patterns between in situ and accreted
stars is a qualitative trend that is robust to the details of the
simulated halos' MDFs.

We predict that a large systematic survey of the detailed chemical
abundances of inner halo stars in the Milky Way will exhibit such dual
trends in [Fe/H] vs [$\alpha$/Fe]. At higher metallicities,
[Fe/H]$\geq -1.0$, Milky Way stars in the inner several kpc of the
halo should exhibit a bimodal distribution in [$\alpha$/Fe]. Unless
the Galaxy has experienced a recent major major, high [$\alpha$/Fe]
stars in this metallicity range will have formed in the Milky Way's
disk or bulge, and then been displaced into halo orbits. Low
[$\alpha$/Fe] stars in this regime will likely have formed in smaller
potential wells than the Milky Way, and were accreted from satellite
companions at high redshift. In fact, such trends in $\alpha$ and Fe
have already been observed in a small sample of 94 local Milky Way
halo dwarfs \citep{Nissen2010}. Larger planned surveys, like HERMES,
which will obtain high resolution spectra of MW halo stars will be
able to observe such patterns, and quantify the importance of
different physical processes to the MW's halo formation.

\section*{Acknowledgments}
We thank the anonymous referee for helping to greatly improve the
paper. We thank Chris Brook for helpful conversations, and Joe Cammisa
at Haverford for computing support. A.Z. and B.W. acknowledge support
from NSF grant AST-0908446. All simulations were run using the NASA
Advanced Supercomputer Pleiades. FG acknowledges support from the HST
GO-1125, NSF AST-0607819 and NASA ATP NNX08AG84G
grants. A.B. acknowledges support from the Sherman Fairchild
Foundation.

\end{document}